\documentclass[twocolumn,preprintnumbers,superscriptaddress,endnote,nofootinbib,prd]{revtex4}

\pdfoutput=1

\usepackage{graphicx}
\usepackage{latexsym}
\usepackage{amsfonts}
\usepackage{amssymb}
\usepackage{amsmath}
\usepackage{slashed}
\usepackage{array}
\usepackage{dcolumn}
\usepackage{feynmp}
\usepackage[hypertex]{hyperref}

\def\lsim{\mathrel{\raise.3ex\hbox{$<$\kern-.75em\lower1ex\hbox{$\sim$}}}}
\def\gsim{\mathrel{\raise.3ex\hbox{$>$\kern-.75em\lower1ex\hbox{$\sim$}}}}

\usepackage{xcolor}
\definecolor{red}{rgb}{1.0, 0, 0}

\newcommand{ \slashchar }[1]{\setbox0=\hbox{$#1$}   
   \dimen0=\wd0                                     
   \setbox1=\hbox{/} \dimen1=\wd1                   
   \ifdim\dimen0>\dimen1                            
      \rlap{\hbox to \dimen0{\hfil/\hfil}}          
      #1                                            
   \else                                            
      \rlap{\hbox to \dimen1{\hfil$#1$\hfil}}       
      /                                             
   \fi}                                             %

\newcommand{\gev}{\text{GeV}}
\newcommand{\tev}{\text{TeV}}
\newcommand{\jets}{\text{jets}}
\newcommand{\pb}{\text{pb}}
\newcommand{\fb}{\text{fb}}

\newcommand{\met}{\slashchar{E}_T}

\begin{document}

\title{Same bump, different channel: Higgs fakes from technicolor }
\author{Adam Martin}
\affiliation{Theoretical Physics Department, Fermi National Accelerator Laboratory, Batavia, IL 60510, USA}
\preprint{FERMILAB-PUB-11-387-T}
\date{\today}

\begin{abstract}
Resonant production of a light, narrow techni-rho, followed by the decay $\rho_T \rightarrow W^{\pm}+\pi_T(jj)$ has been proposed as an explanation of the $W+jj$ excess observed by CDF. If the $\pi_T$ decays to $\tau+\nu_{\tau}/\tau^+ \tau^-$ rather than to jets, subsequent leptonic $\tau$ decay leads to $\ell^+\ell^- \nu\nu\bar{\nu}\bar{\nu}$, a final state that will be picked up by standard model $WW^{(*)} \rightarrow \ell^+\ell^- +\met$ Higgs searches. We point out that, for the same range of technicolor parameters required to fit the CDF $W+jj$ excess, the correlated $\ell^+\ell^- + \met$ technicolor signal can have strength comparable to that of an intermediate-mass standard model Higgs boson, and therefore could be visible at the Tevatron or LHC. 
\end{abstract}

\maketitle

\section{Introduction \label{sec:intro}}
In the range of masses $m_H \sim 130 - 180\,\gev$, the most sensitive Higgs search mode at the LHC and Tevatron is $h \rightarrow WW^{(*)} \rightarrow \ell^+\ell^-\nu\bar{\nu}$~\cite{Aad:2011qi, ATLHIGGS, Chatrchyan:2011tz,CDF:2011cb}.
As there are two sources of missing energy, the invariant mass of the Higgs cannot be reconstructed. Thus, any new physics with a  $\ell^+\ell^- + \met$ final state, regardless of whether the leptons and missing energy came from $W$ bosons, will be picked up to some extent by Higgs searches.  This makes the $\ell^+\ell^- + \met$ mode particularly susceptible to `Higgs-fakes' -- non-Higgs particles which get mistaken for the Higgs. 

While there are many potential Higgs-fakes, in this paper we focus on fake Higgs signals from technicolor, namely $pp \rightarrow \rho_T \rightarrow W(\ell \nu)\pi_T$ where $\rho_T$ is a techni-rho and $\pi_T$ is a techni-pion. Techni-pion couplings to electrons and muons are usually negligible, however the coupling, and hence the branching ratio, of $\pi_T$ to $\tau + \nu_{\tau}$ can be large. A leptonic decay of the tau then gives us the $\ell^+\ell^-\nu\nu\bar{\nu}\bar{\nu}\,\cong \ell^+\ell^- + \met$ signature necessary for posing as a Higgs boson.  The technicolor signal is controlled by a handful of parameters: $M_{\rho_T}, M_{\pi_T}$ govern the rate for $\rho_T \rightarrow W\pi_T$ production, while $B(\pi_T \rightarrow \tau\nu_{\tau}/\tau^+\tau^-)$  determines the fraction of $\ell^+\ell^-+\met$ final states.

The technicolor Higgs-fake is interesting for several reasons. First, one (or both) of the leptons come from a $\tau$ rather than a $W$ and there are four invisible particles rather than two. This makes the $W(\ell\nu)\pi_T(\tau\nu)$ signal qualitatively different from Higgs impostors considered in the past~\cite{Higgsimpostors}. Second, when a techni-pion decays hadronically, the final state from  $\rho_T \rightarrow W + \pi_T$ is a $W$ and two jets.  This final state has been the subject of much interest since the CDF collaboration reported an excess in $W(\ell\nu) + jj$ of $3.2\,\sigma$~\cite{Aaltonen:2011mk} (later updated to $4.1\,\sigma$), and in Ref~\cite{Eichten:2011sh} it was shown that, for $M_{\rho_T} \sim 250-300\,\gev$, $M_{\pi_T} \sim 150-170\,\gev$, a technicolor $\rho_T \rightarrow W + \pi_T$ explanation fits the excess quite well. 

As we will show here, for the same range of parameters required to fit the CDF excess, the $\ell^+\ell^-+\met$ signal from the techni-pion + W channel can have a rate similar to an intermediate mass standard model (SM) $h\rightarrow WW^{(*)} \rightarrow \ell^+\ell^-\nu\bar{\nu}$.
Thus, the {\em same} (non-Higgs) physics that leads to $W+jj$ can also generate a $\ell^+\ell^-+\met$ signal that fakes a Higgs; given the sensitivity of Tevatron and LHC searches, such an excess should be visible, either now or in the near future.
This coincidence of signals is especially tantalizing given that both CMS and ATLAS see a discrepancy between their observed and expected limits in the $\ell^+\ell^-+\met$ channel for $~120\,\gev \lesssim m_H \lesssim 180\,\gev$~\cite{HiggsEPS}. 

The setup of this paper is the following: after giving an introduction to the interactions and states in technicolor in Sec.~\ref{sec:techni}, we describe the CDF $W+jj$ signal in Sec.~\ref{sec:wjj} along with the technicolor parameters which fit the observed excess. Then, in Sec.~\ref{sec:higgspose} we study the technicolor $\ell^+\ell^-+\met$ signal: its rate, its kinematics, and its dependence on model parameters. We also plot several important kinematic variables and compare with SM Higgs signals. Finally, in Sec.~\ref{sec:conclu}, we conclude.

\section{Techni-pions and Techni-rhos \label{sec:techni}}

Techni-pions and techni-rhos are common ingredients in theories of dynamical electroweak symmetry breaking. Techni-pions are pseudo-scalar states, uneaten cousins of the longitudinally polarized $W/Z$; they are present only if the chiral symmetry of the underlying degrees of freedom is big enough. Being pseudo-Goldstone bosons, techni-pions are typically the lightest states in the spectrum, and are expected to have small couplings to light fermions. Techni-rhos are spin-1 resonances which couple strongly to (longitudinally polarized) $W/Z$ and techni-pions, but have very weak couplings to SM fermions. Both techni-pions and techni-rhos lie in electroweak triplet representations.

Given their connection to electroweak symmetry breaking, the most natural mass for techni-resonances is $\mathcal O(\tev)$, far too heavy to be produced with sizable rate at the Tevatron. One class of models with lighter techni-resonances, and thus a larger rate, are `multi-scale' technicolor models~\cite{Lane:1989ej}, where there is more than one (presumably dynamical) source of electroweak breaking. In the simplest version, there are two scales (or vacuum expectation values), $v_1,\, v_2$. The sum in quadrature of the two scales is fixed to the weak scale, but the ratio of the vacuum expectation values is a free parameter. Resonances associated with the lighter scale are not only light, but their couplings to SM fermions are also suppressed~\cite{Lane:2009ct}:
\begin{equation}
g_{ff\rho_{T,1}} \sim g \Big( \frac{M_W}{M_{\rho_{T,1}}} \Big) \sin{\beta},
\end{equation}
where $\tan{\beta} = v_1/v_2$ and $g$ is a SM electroweak gauge coupling. Provided $g_{ff\rho_{T,1}} \lesssim 0.1\times g$, resonances as light as $200-300\,\gev$ are allowed by all LEP, Tevatron, and LHC data. While resonance-fermion couplings are small in these scenarios, the resonance-$\pi_T$ couplings are large; as the physical (uneaten) combination of $\pi_T$ in a multi-scale model reside primarily in the sector with the smaller contribution to EWSB, the $\pi_T$ couple more strongly to the lower-scale resonances.

\section{CDF $W$ + dijet excess \label{sec:wjj}}
 
 A two-scale technicolor setup was invoked in Ref.~\cite{Eichten:2011sh} to explain the $3.2\,\sigma$ excess in $W+jj$ events observed by the CDF collaboration in $4.3\,\fb^{-1}$ of data~\cite{Aaltonen:2011mk}\footnote{After $7.3\,\fb^{-1}$ the significance grew to $4.1\,\sigma$. The update, many cross-checks, and several kinematic distributions can be found in Ref.~\cite{CDFWjjwebsite}.}.
In the CDF signal data, the invariant mass of the dijets, after subtracting the large $W+\jets$ background, forms a peak centered at $\sim150\,\gev$. Assuming the extra events come from new physics, and assuming the signal acceptance is similar to $p\bar p \rightarrow H+W$, the cross section attributed to this excess is $3.1\pm 0.8\,\pb$~\cite{ViviEPS}. The D\O\ collaboration explored the same final state using identical cuts in Ref.~\cite{Abazov:2011af}, setting a limit on the allowed new physics cross section of $0.82^{+0.83}_{-0.82}\,\pb$, consistent with no excess. While the cuts are the same in both collaborations, the treatment of systematics and, in particular, the jet-energy correction are quite different~\cite{Buckley:2011hi,CDFWjjwebsite, ViviEPS}. Given the large background that must be subtracted to see the signal, all of these subtle differences must be examined carefully\footnote{Note the D\O\ analysis utilized less data than the updated CDF analysis and did not contain an inclusive ($\ge 2\, \jets$) analysis, which would ameliorate the differences in jet definition between collaborations}. 
The LHC will eventually weigh in on this issue. However, with the current amount of integrated luminosity, ATLAS/CMS are not yet sensitive to weakly coupled, $q-\bar q$ initiated processes such as techi-rho production~\cite{EPSATLAS, Eichten:2011xd, Harigaya:2011ww, Buckley:2011hi}.

To fit the CDF data, in Ref.~\cite{Eichten:2011sh} the parameters $M_{\rho_T} = 290\,\gev, M_{\pi_T} = 160\,\gev, \sin{\beta} = 1/3$ were used. Not only does the $W+jj$ rate for this parameter set match the excess observed by CDF, but the signal changes with respect to the background as cuts are varied in exactly the way the technicolor model predicts~\cite{CDFWjjwebsite, ViviEPS}. Furthermore, studies of the kinematic distributions of the CDF excess are consistent with technicolor expectations~\cite{Buckley:2011hi}. Specifically, the dijet $\Delta R$ distribution has a sharp turn on at $\sim 2.3$ and the $p_T$ of the dijet system falls to zero above $\sim 80\,\gev$~\cite{LANE}. Both of these features are expected in a two-resonance topology, where the total energy in the $W+$ dijet system is limited by parent resonance mass\footnote{The numerical values where the $p_T$ and $\Delta R_{jj}$ features occur depends sensitively on the mass difference between the parent and daughter resonances}. Another interesting aspect of the parameters used in Ref.~\cite{Lane:1989ej, Eichten:2011sh}  is that the techni-rho mass is less than twice the techni-pion mass. This choice kinematically suppresses the $\rho_T \rightarrow \pi_T \pi_T$ decay mode, leaving $\pi_T + W/Z$ as the dominant mode. The dominance of the  $\pi_T + W/Z$ mode, along with the small width it induces, are the keys to a large technicolor rate into $W + jj$.

While the point $M_{\rho_T} = 290\,\gev, M_{\pi_T} =160\,\gev$ was chosen for calculations in~\cite{Eichten:2011sh}, the actual range of masses which fit the signal is broader, approximately $M_{\rho_T} \sim 250-300\,\gev, M_{\pi_T} \sim 150-170\,\gev$. As the $\rho_T$ is taken lighter, the $W\pi_T$ rate increases. However, an increasing cross section can be counteracted by decreasing the $M_{\rho_T} - M_{\pi_T}$ splitting; as the available kinetic energy for the $W-\pi_T$ system shrinks, the $p_{T,jj}$ cut imposed by CDF~\cite{Aaltonen:2011mk} removes more and more of the $W+jj$ signal.

In Ref.~\cite{Eichten:2011sh}, the flavor of the jets coming from charged $\pi_T$ was left ambiguous. Naively, techni-pions couple to SM fermions proportional to their mass. With the techni-pion mass less than the top mass, $\pi^{+} \rightarrow \bar b c$ is an obvious choice for  the dominant mode. However, the flavor structure of the techni-pion-SM fermion coupling is model dependent, so inter-generation decays can be accompanied by CKM-like mixing factors. Taking this mixing as an input parameter, we can freely dial the branching fraction of $\pi_T$ to bottom quarks. If decays to bottom + $u/c$ are highly suppressed, the most massive fermion to decay into is the $\tau$. If the $\pi_T$ decays appreciably to $\tau + \nu_{\tau}$, the rate for $W+jj$ suffers, however, the $\tau$ modes are essential for faking a Higgs signal. 

Charged techni-pion decays are not the only way to get dilepton states from technicolor. As the $\pi^0_T \rightarrow \bar b b$ mode is kinematically allowed and is not an inter-generation decay, it is not expected to be suppressed with mixing angles so we will leave the $\pi^0_T$ branching fractions unchanged. While $B(\pi^0_T \rightarrow \bar b b)$ is the dominant decay mode, $B(\pi^0_T \rightarrow \tau^+\tau^-)$ is not negligible ($\sim 10\%$) and can still lead to $\ell^+\ell^- +\met$ final states -- either from both taus decaying leptonically or one leptonic tau plus a lepton from the accompanying $W$. However, $\ell^+\ell^- +\met$ events from $W+\pi^0_T$ decays usually contain extra jets, and Higgs searches often include a jet veto to remove top-quark background. Under a jet veto, only the fraction of  $W+\pi^0_T(\tau\tau)$ events where all jets are too soft or fall outside the detector acceptance will actually survive. 

\section{Higgs posing \label{sec:higgspose}}

In this section we will show how related decay modes of the $\pi_T$ can lead to a Higgs-fake $\ell^+\ell^-+\met$ signal and explore the rates and kinematics of such signals. An example Feynman diagram for the Higgs-faking processes we have in mind is shown below in Figure~\ref{fig:llnunu}.
\begin{figure}[!h]
\centering
\includegraphics[width=3.0in, height =1.75in]{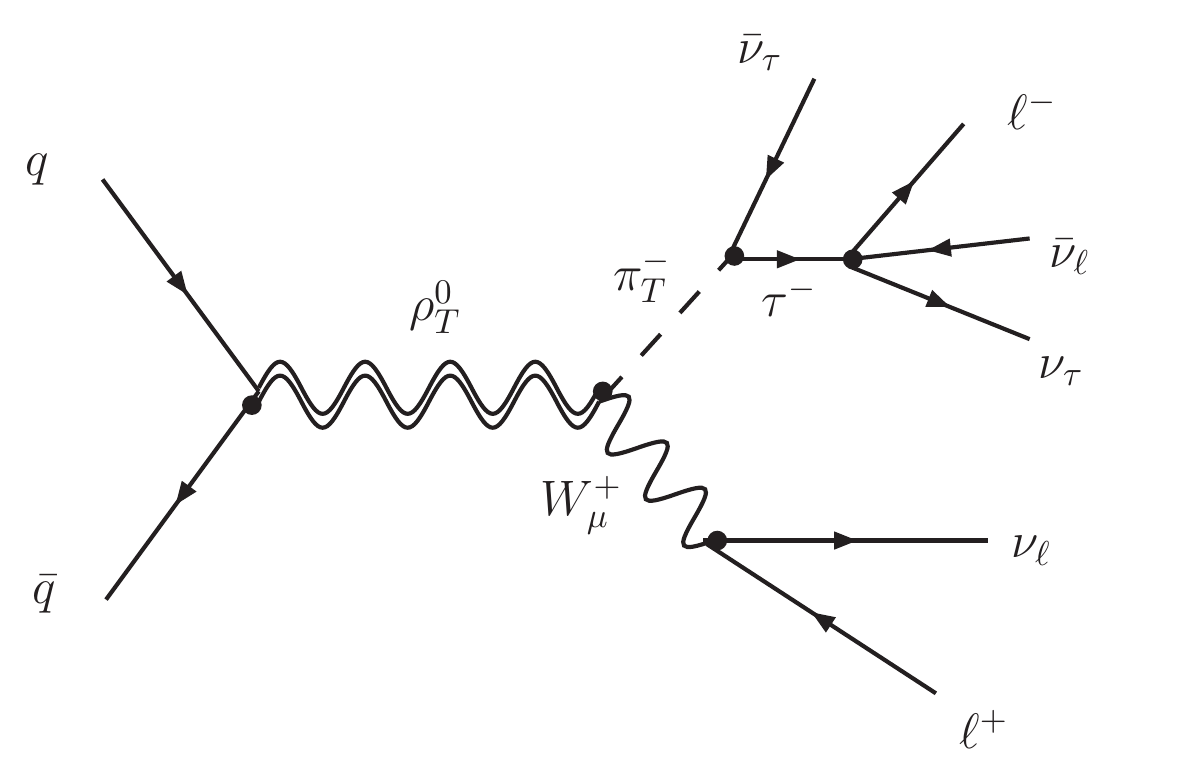}
\caption{Diagram for the process $pp \rightarrow \rho_T \rightarrow W^{\pm}(\ell\,\nu)\,\pi^{\mp}_T(\tau^{\mp\,(}\overline{\nu}^{)})$. }
\label{fig:llnunu}
\end{figure}

To generate a dilepton + $\met$ final state from the technicolor scenario, we must pay the price of $B(\pi_T \rightarrow \tau\nu)$ and the $\sim 35$\% leptonic branching fraction of the tau. Additionally, the Higgs-fake events must pass selection cuts. On top of basic identification and kinematic cuts, $\ell^+\ell^-+\met$ Higgs searches impose low dilepton invariant mass, substantial dilepton $p_T$ and a small azimuthal separation between leptons~\cite{ATLHIGGS}. The first two cuts are imposed to remove background events containing $Z$ bosons, while the third cut takes advantage of the fact that the two leptons in the (Higgs) signal originate in a spin-0 resonance. While the technicolor fake signal has no $Z$, the mass of the mother resonance ($M_{\rho_T} \gtrsim 250\,\gev$) is considerably higher than that of an intermediate-mass Higgs, which raises the $m_{\ell\ell}$ distribution to higher values. The techni-signal also has a different correlation among leptons than a Higgs, so the efficiency  for the $\Delta \phi_{\ell\ell}$ cut can be quite different.

To study the relative efficiencies under Higgs cuts, we rely on Monte Carlo. Technicolor events are generated using PYTHIA6.4~\cite{Sjostrand:2006za}, while comparison samples of $h\rightarrow WW^{(*)} \rightarrow \ell^+\ell^-\nu\bar{\nu}$ events are generated at matrix-element level with MadGraph5~\cite{Alwall:2011uj}, then passed through PYTHIA for showering and hadronization. Post-PYTHIA, isolated leptons and photons are identified and removed from the list of final particles\footnote{Leptons are considered as isolated if the total $E_T$ of all particles within a radius of $R=0.4$ of the lepton is less that $0.2\times E_{T, \ell}$.}. The remaining particles are granularized into $0.1\times 0.1$ cells in $(\eta, \phi)$ space, with the energy of each cell rescaled such that it is massless. All cells with energy greater than $1.0\,\gev$ are then clustered into jets using the anti-$k_T$ algorithm via FastJet~\cite{Cacciari:2005hq}. 

Once all physics objects have been identified, we apply the (below-threshold) Higgs search cuts~\cite{ATLHIGGS}. The basic cuts are the following:  two leptons $p_{T, \ell1} > 25.0\,\gev,\, p_{T, \ell 2} > 20.0\,\gev,\, |\eta_{\ell}| < 2.5,\, \met > 25\,\gev,\, 10\,\gev < m_{\ell\ell}$ and $|\,m_{\ell\ell} - M_Z\,| > 15\,\gev$ for same-flavor lepton pairs.  On top of the basic cuts, topological cuts $p_{T, \ell\ell} > 30.0,\, m_{\ell\ell} < 50.0\,\gev,\, \Delta \phi_{\ell\ell} < 1.3$ are applied\footnote{Technically, the cuts in~\cite{ATLHIGGS} depend on the flavor of the leptons. While the cuts we describe in the text are the $e\mu$ cuts, in our simulations we vary the cuts with lepton flavor as dictated by Ref.~\cite{ATLHIGGS}.}. Finally, events are binned by jet multiplicity ($p_{T,j} > 25\,\gev, |\eta_j| < 4.5$). We will focus on the zero-jet bin, which has the best sensitivity. To get an idea for how the $m_{\ell\ell}$ and $\Delta \phi_{\ell\ell} $ cuts effect the technicolor and SM Higgs signals differently, we plot the shapes of these distributions below in Fig.~\ref{fig:distributions}. The plots show two different technicolor benchmark ($M_{\rho_T}, M_{\pi_T})$ points, as well as two different SM Higgs masses $m_H = 130\,\gev$ and $m_H = 150\,\gev$.
\begin{figure}
\centering
\includegraphics[width=3.0in, height = 2.0in]{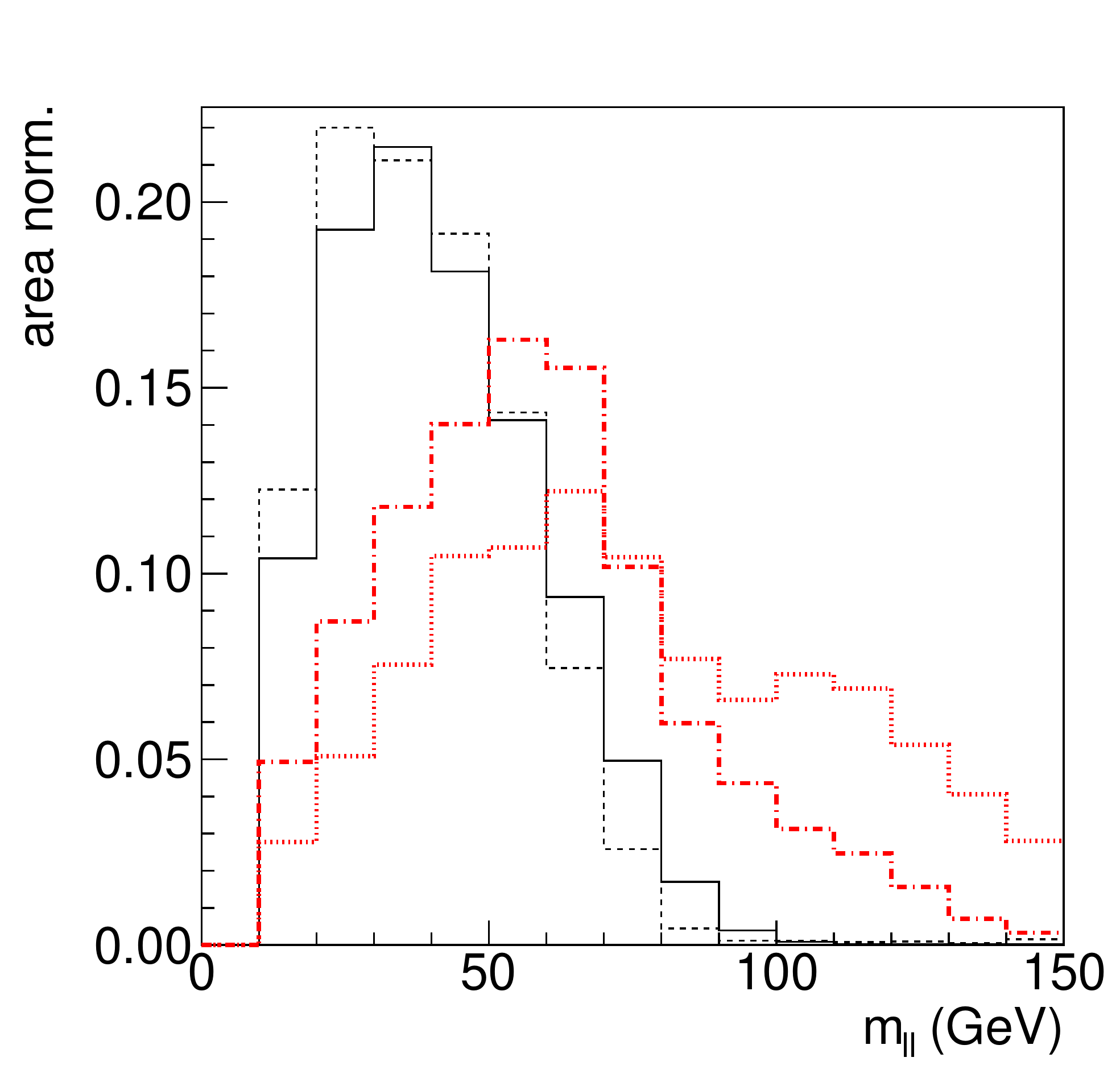} \\
\includegraphics[width=3.0in, height = 2.0in]{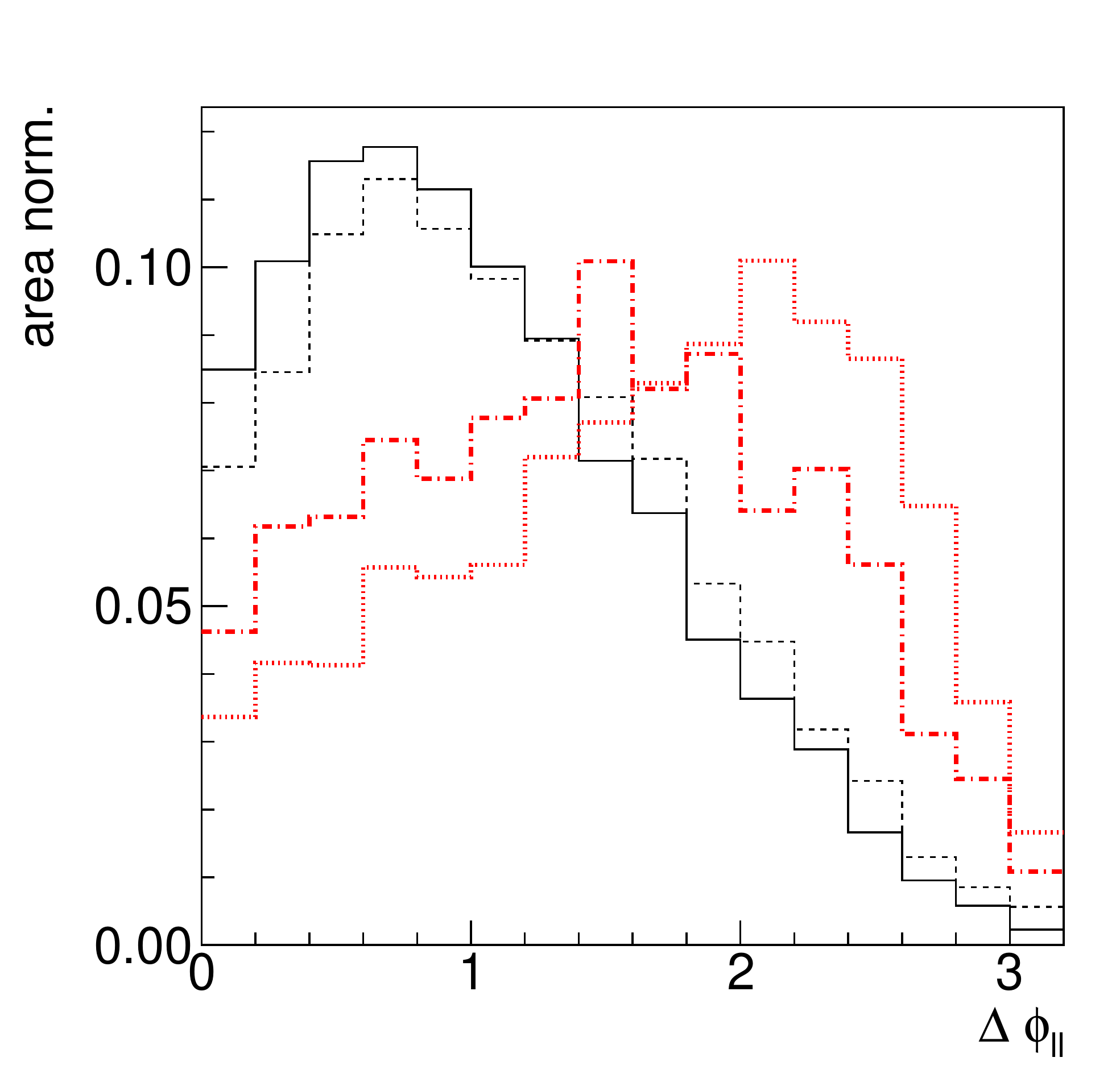} 
\caption{Top panel: $m_{\ell\ell}$ distributions for two SM Higgs mass values, $m_H = 150\,\gev$ (black, solid), $m_H = 130\,\gev$ (black, dashed), and for two different technicolor benchmark scenarios, $M_{\rho_T} = 290\,\gev, M_{\pi_T} = 160\,\gev$ (red, dot-dashed) and $M_{\rho_T} = 250\,\gev, M_{\pi_T} = 150\,\gev$ (red, dotted). Only basic selection cuts have been applied. Bottom panel: $\Delta \phi_{\ell\ell}$ distribution, after basic selection, for the same SM and technicolor points. All distributions are normalized to unit area.}
\label{fig:distributions}
\end{figure}
The technicolor signal clearly has a different shape than a SM Higgs, peaking at higher $m_{\ell\ell}$ and $\Delta \phi_{\ell\ell}$. The technicolor rates would certainly grow if the $m_{\ell\ell}, \Delta \phi_{jj}$ cuts were loosened, however, a larger acceptance can be compensated for by taking a smaller $B(\pi_T \rightarrow \tau \nu_{\tau})$.

\begin{figure*}
\centering
\includegraphics[width=2.1in, height=2.35in]{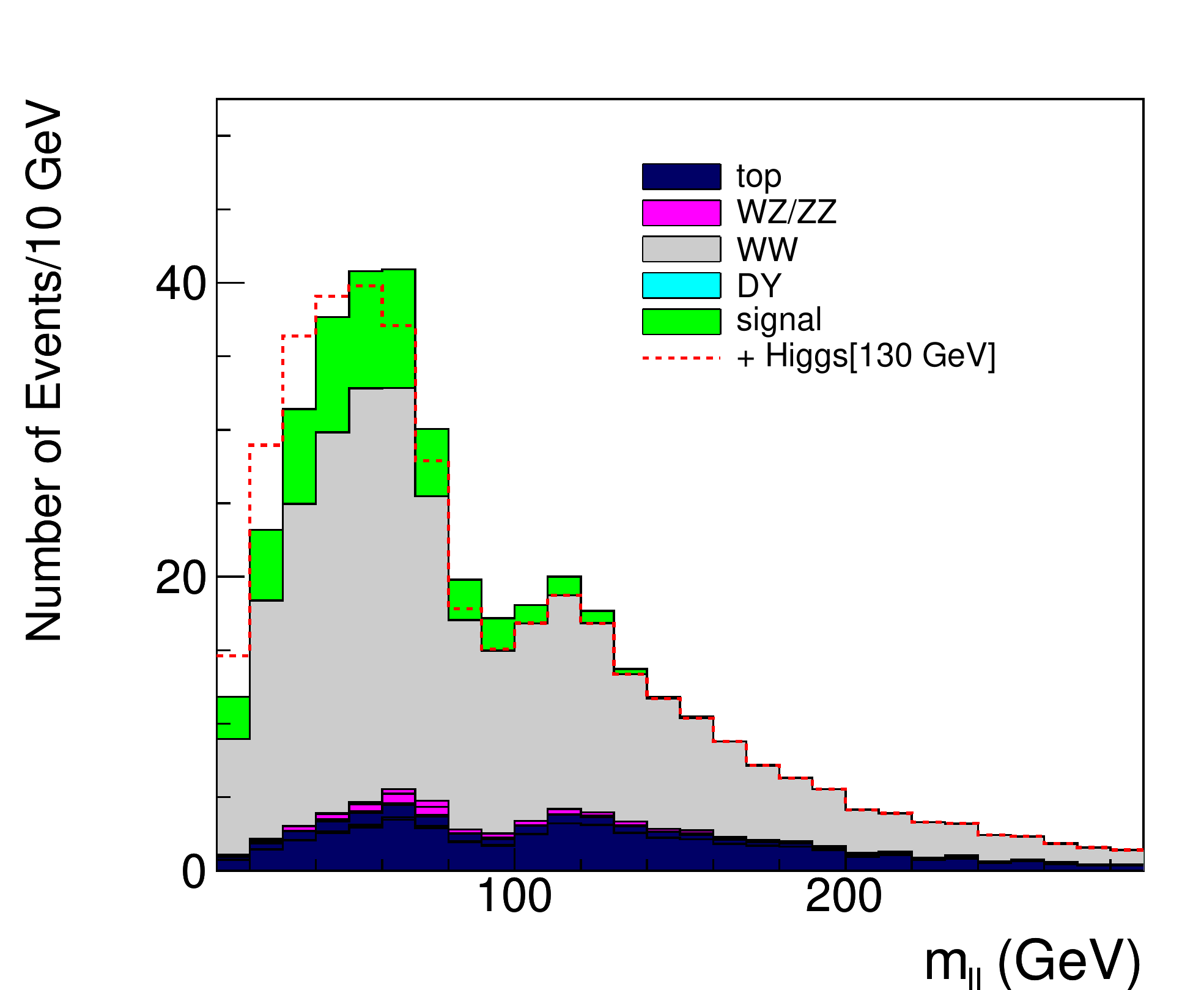}\hspace{0.2in}
\includegraphics[width=2.1in, height=2.35in]{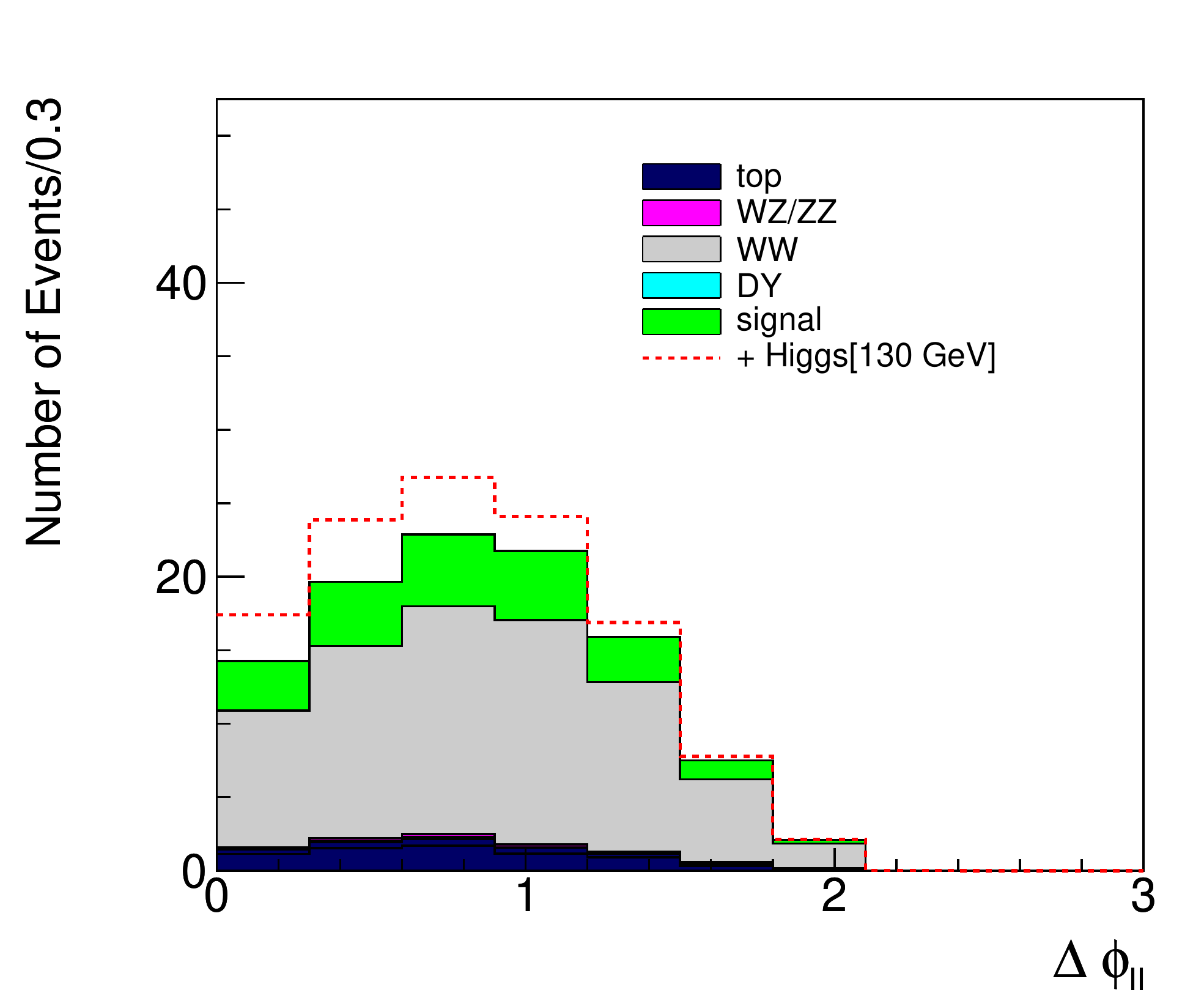}\hspace{0.2in}
\includegraphics[width=2.1in, height=2.35in]{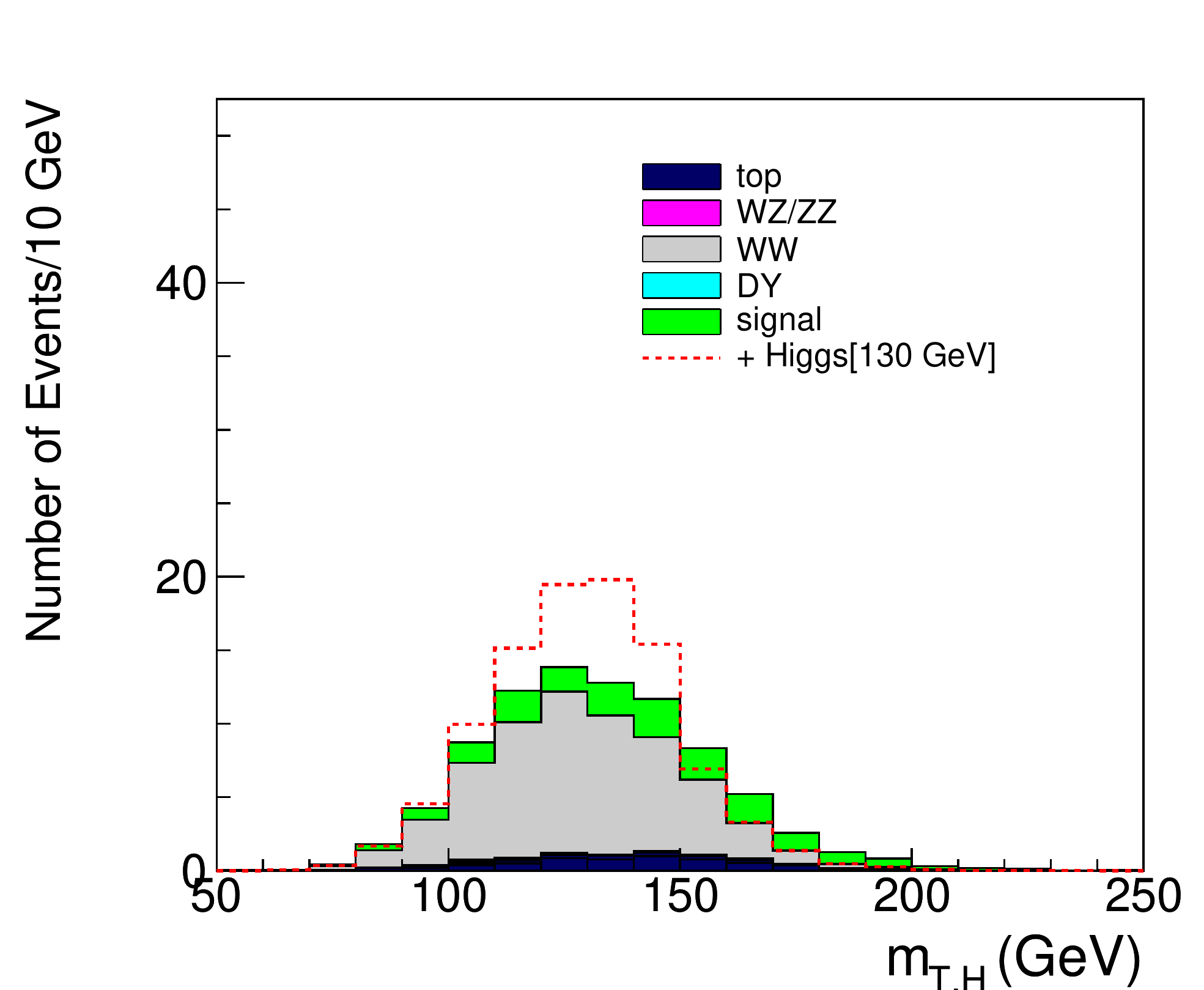}
\caption{Comparison of SM background plus technicolor signal $(M_{\rho_T} = 250\,\gev, M_{\pi_T} = 150\,\gev)$ (histogram) with the background plus SM Higgs of mass $m_H = 150\,\gev$ (dashed line): $m_{\ell\ell}$ distribution after basic cuts and the $p_{T, \ell\ell}$ cut (left pane), $\Delta \phi_{\ell\ell}$ distribution after basic cuts and cuts on the dilepton $p_T$ and invariant mass (center pane), and $m_{T,H}$ after all cuts (right pane). All plots assume $1\,\fb^{-1}$ of data at a $7\,\tev$ LHC.}
\label{fig:mh150_compare}
\end{figure*}

\begin{figure*}
\centering
\includegraphics[width=2.1in, height=2.35in]{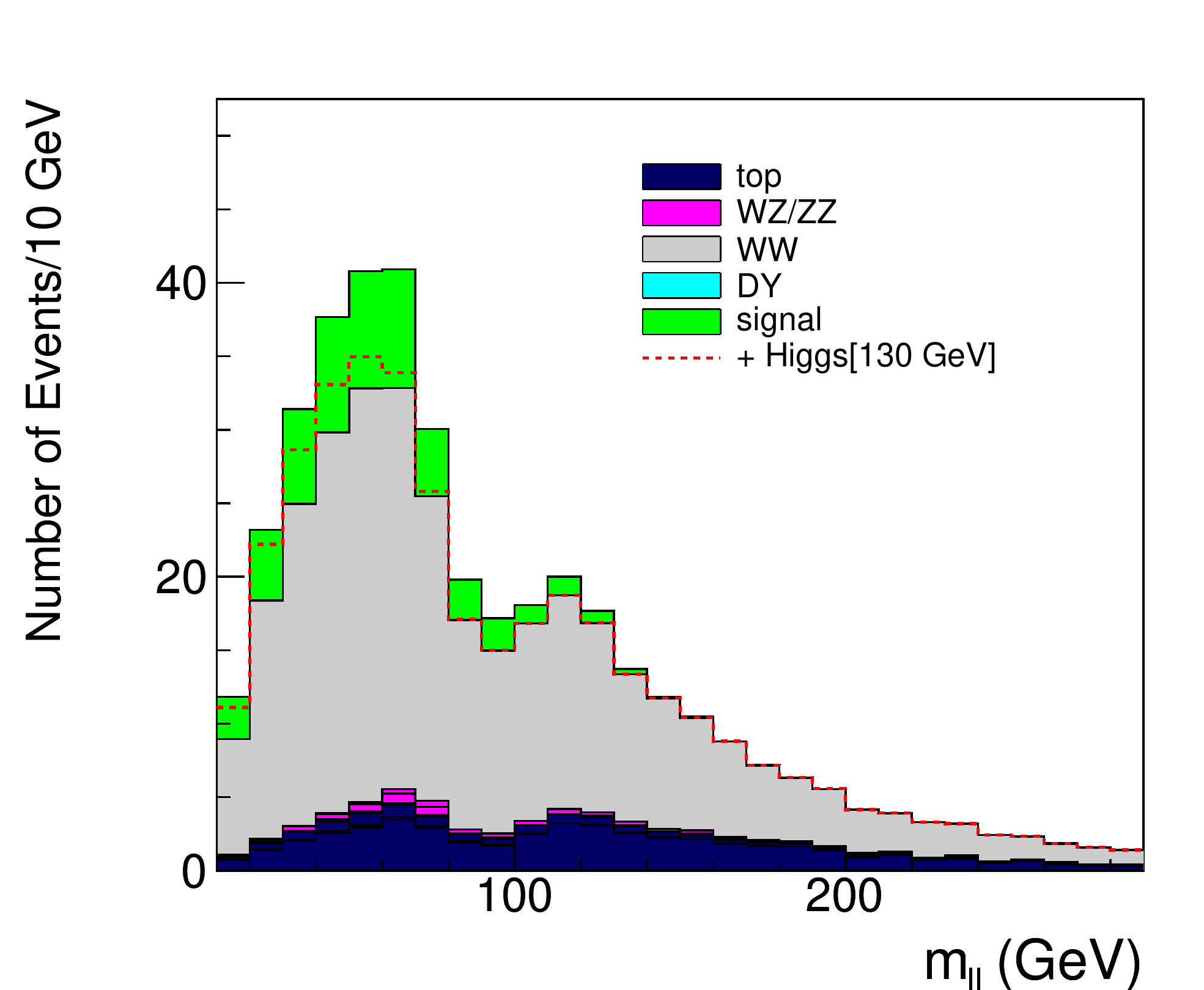}\hspace{0.2in}
\includegraphics[width=2.1in, height=2.35in]{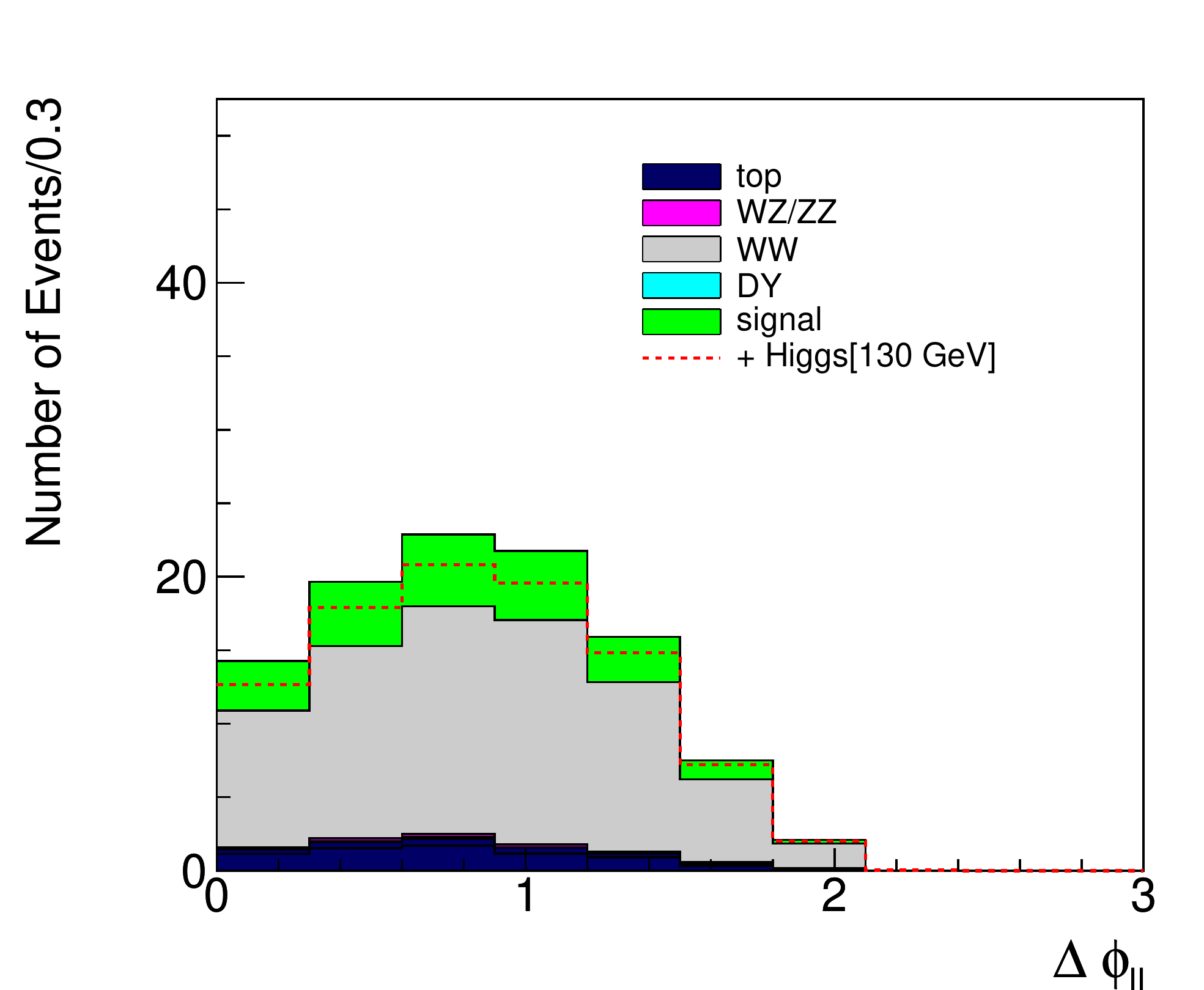}\hspace{0.2in}
\includegraphics[width=2.1in, height=2.35in]{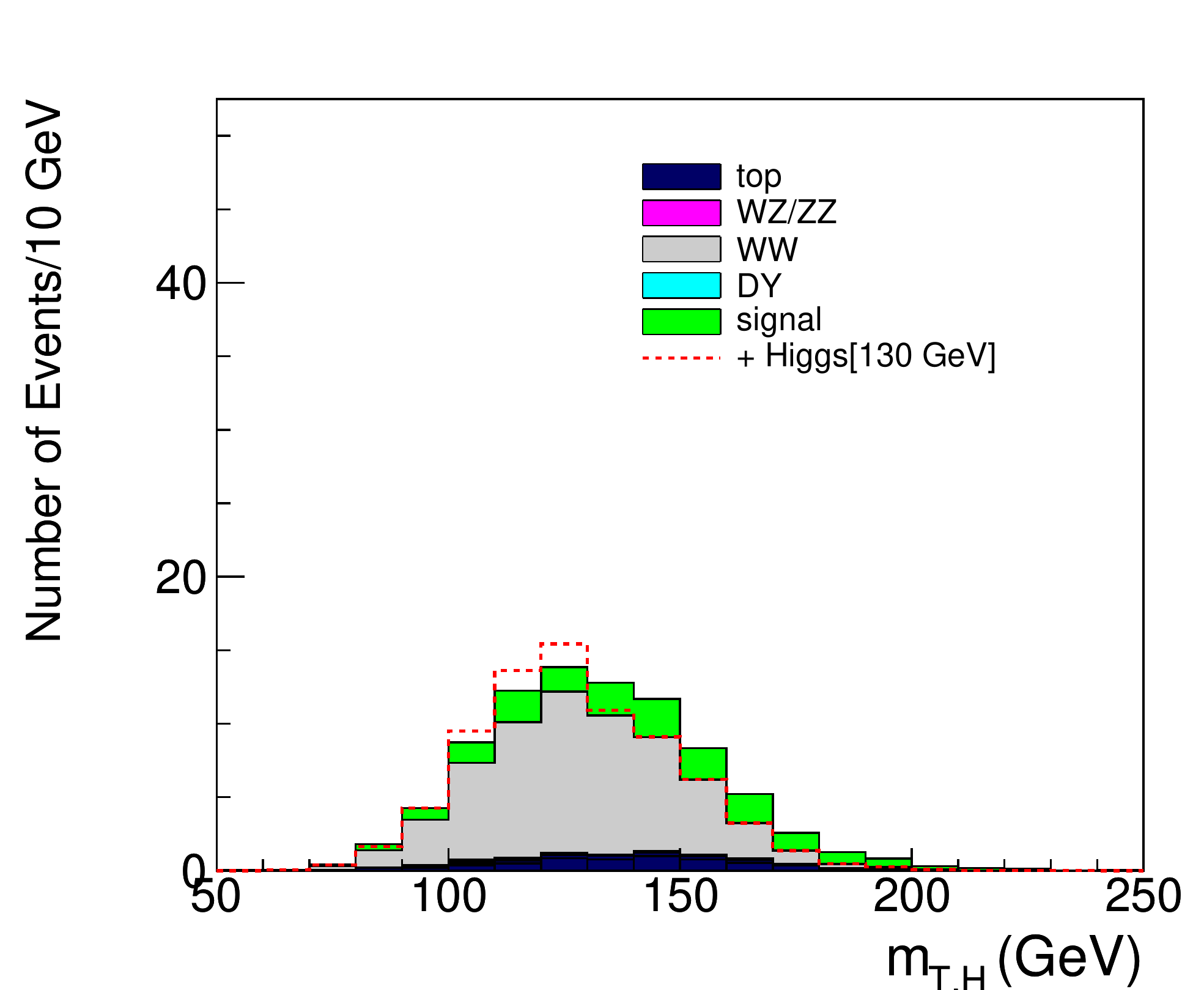}
\caption{Comparison of SM background plus technicolor signal $(M_{\rho_T} = 250\,\gev, M_{\pi_T} = 150\,\gev)$ (histogram) with the background plus SM Higgs of mass $m_H = 130\,\gev$ (dashed line). The same cuts have been imposed as in Figure~\ref{fig:mh130_compare}.}
\label{fig:mh130_compare}
\end{figure*}

After imposing the $p_{T,\ell\ell},\,m_{\ell\ell},\,\text{and}\, \Delta \phi_{jj}$ cuts, the final cross sections for the two technicolor mass points are tabulated below in Table~\ref{table:comparison} and compared with the two SM Higgs masses used in Figure.~\ref{fig:distributions}. As we are seeking to link the CDF $W+jj $ excess with a possible $\ell^+\ell^-+\met$ excess, in Table~\ref{table:comparison} we also present the cross sections (at the Tevatron) for the technicolor points after imposing the $W+jj$ cuts from Ref.~\cite{Aaltonen:2011mk}. Also, for each technicolor point we present two cross sections; the first number uses $B(\pi_T \rightarrow \tau \nu_{\tau}) = 4.5\%$ -- the PYTHIA default inspired by Ref.~\cite{Lane:1999uh,Lane:2002sm} which assumes no suppression of the $\pi_T \rightarrow b c/b u$ or $d c$ modes, while the second number uses $B(\pi_T \rightarrow \tau \nu_{\tau}) = 66\%$, the result of shutting off all inter-generational $\pi_T$ decays. These two numbers demonstrate how the technicolor rates depend on the branching fraction of the charged $\pi_T$ to $\tau\nu_{\tau}$. All technicolor cross sections include a $K$-factor of $1.3$~\cite{Lane:1999uh, Lane:1999uk} to account for NLO effects, and we have scaled the SM Higgs cross sections to match the NLO values computed with MCFM~\cite{Campbell:2011bn}. 

\begin{table}[h!]
\begin{tabular}{|c|c|c|}\hline
Process & $\sigma(\ell^+\ell^-\nu\bar{\nu})_{LHC7}$& $\sigma(\ell\,\nu\,j\,j)_{\text{TeV}}$\\ \hline
$m_H= 130\,\gev$ & $10\, \fb$ & -- \\ \hline
$m_H = 150\,\gev$ & $33\,\fb$& -- \\ \hline \hline
$M_{\rho_T} = 290\,\gev$ & $0.81\,\fb$& 88 fb \\
~~$M_{\pi_T} = 160\,\gev$ & $9.1\,\fb$ & 53 fb \\ \hline
$M_{\rho_T} = 250\,\gev$ & $2.7\,\fb$& 82 fb \\
~~$M_{\pi_T} = 150\,\gev$ &  $19\,\fb$& 58 fb \\ \hline
\end{tabular}
\caption{Cross sections, after all branching fractions and cuts. The second column shows $\sigma(pp \rightarrow \ell^+\ell^-+\met)$ at the LHC, while the third column shows $\sigma(p\bar p \rightarrow \ell\nu\, jj)$ at the Tevatron. For the technicolor benchmarks, the first row assumes $B(\pi^{\pm} \rightarrow \tau \nu_{\tau}) = 4.5\%$ (PYTHIA default), while the second row assumes inter-generational $\pi^{\pm}_T$ decays are impossible and hence $B(\pi^{\pm} \rightarrow \tau \nu_{\tau}) = 66\%$. The technicolor cross sections include both $W\pi^0_T$ and $W\pi^{\pm}_T$ production.}
\label{table:comparison}
\end{table}

No jet/lepton smearing or detector inefficiencies have been taken into account in these simulations. While these detector effects will influence the total rates, they will effect Higgs and technicolor impostor signals in the same way, so the relative rates in the second column of Table~\ref{table:comparison} will remain unchanged. Note that, while the branching fraction for $\pi^{\pm}_T$ changes by a factor of $12$ between the two rows (in each technicolor point), the actual changes in the $\ell^+\ell^-+\met$ rates are less drastic. The mismatch occurs because the rows differ only in the charged $\pi_T$ contribution to $\ell^+\ell^-+\met$ while the $\pi^0_T$ contribution is held fixed. 

For the technicolor parameter point used in Ref.~\cite{Eichten:2011sh}, the rates in $\ell^+\ell^-+\met$ are smaller than the rate for either SM Higgs values regardless of the $\pi^{\pm}_T$ branching fraction to taus. The technicolor $\ell^+\ell^-+\met$ signal in this case would look like a (non-SM) Higgs with decreased couplings. For the lighter mass technicolor point, the $\ell^+\ell^-+\met$ rate is much closer to the SM Higgs rate. Depending on $B(\pi^{\pm}\rightarrow \tau\nu)$, the rate can even be larger than a SM Higgs of mass $130\,\gev$. To determine whether a technicolor point fits the CDF $W+jj$, recall that the excess consisted of $\sim 255 \pm 58$ events in $4.3\,\fb^{-1}$ of luminosity~\cite{Aaltonen:2011mk}. Looking at the third column, we can see that the lighter technicolor point is still quite compatible.

To further illustrate how the technicolor Higgs-fake signal would look compared to a SM Higgs, we plot some of the important distributions, below in Figures~\ref{fig:mh150_compare}, \ref{fig:mh130_compare}. In these comparison plots, the technicolor signal is stacked on top of SM backgrounds that has been generated using ALPGENv214~\cite{Mangano:2002ea} and PYTHIA\footnote{We use CTEQ6L1 parton distribution functions and take the default factorization/renormalization scale in all simulations}. The dominant background is $WW$, with small contributions from other diboson processes and $\bar t t$. The signal + background distributions we plot are the same as can be found in Ref.~\cite{ATLHIGGS} : i.) $m_{\ell\ell}$ after all basic cuts and the $p_{T,\ell\ell}$ cut, ii.) $\Delta \phi_{\ell\ell}$ after basic cuts, the $p_{T,\ell\ell}$ and $m_{\ell\ell}$ cuts, iii.) the total $\ell^+\ell^-+\met$ transverse mass $m_{T,H}$ after all cuts, where $m_{T,H}$ is defined as

\begin{align}
& m^2_{T,H} = (E_{T, \ell\ell} + \met)^2 - (\vec p_{T, \ell\ell} + \slashchar{ \vec p}_T)^2,\nonumber \\
 &~~E_{T, \ell\ell}  = \sqrt{m^2_{\ell\ell} + p^2_{T, \ell\ell}}.
\end{align}
For leptons coming from a SM Higgs, this distribution has a Jacobian edge at $m_H$, while the technicolor distribution is much wider. Detector effects will smooth out the edge in the Higgs $m_{T,H}$ distribution, but the shape will remain different than we get from technicolor.

In Figures~\ref{fig:mh150_compare}, \ref{fig:mh130_compare} we use the lighter technicolor point throughout and assume the maximum $\pi^{\pm}_T$ branching fraction to tau. In the first (second) row of plots the technicolor signal is compared with a $m_H = 150\,\gev$ $(m_H = 130\,\gev)$ SM Higgs. 
While there are some differences between the technicolor Higgs-fake signal and a genuine Higgs signal, the differences disappear as more cuts are applied. After all selection and topological cuts, the technicolor Higgs-fake is nearly indistinguishable from a $130\,\gev$ SM Higgs, especially with the currently available statistics.

While linking two excess is obviously the most exciting outcome, our result is interesting even if one of these excesses disappears with more data. The lack of a  $\ell^+\ell^-+\met$ signal is informative because it imposes an upper limit on the branching fraction of the $\pi^{\pm}_T$ to $\tau$. Similarly, if the $W+jj$ excess disappears we are free to consider a wider range of technicolor parameters for Higgs-fake signals. Finally, while we have focused on a technicolor explanation of both the $W+jj$ excess and a potential $\ell^+\ell^-+\met$ signal, two-Higgs-doublet-motivated scenarios for $W+jj$, such as~\cite{Cao:2011yt, Chen:2011wp, Fan:2011vw} can also generate Higgs fakes via $W(\ell\nu)+H^{\pm}(\tau\nu)$. Note that two-Higgs doublet scenarios also contain a real $h\rightarrow WW^{(*)} \rightarrow \ell^+\ell^- + \met$ signal, and the interplay of the real and fake Higgs signals would be interesting to study.

\section{Conclusions \label{sec:conclu}}

We have shown that the same dynamics behind CDF's $W+jj$ excess, namely low-scale technicolor, 
 $\ell^+\ell^-+\met$ final states of the same order of magnitude as an intermediate mass SM Higgs. In low-scale technicolor, the $\ell^+\ell^-+\met$ final states come from $W+\pi_T$, where the $\pi_T$ decays to $\tau+\nu /\tau^+\tau^-$. The rate for this process is sensitive to the branching fraction of techni-pions to tau states, which in turn depends on the details of how techni-pions and SM fermions interact.  Varying the charged techni-pion branching fraction to $\tau\nu$ over a reasonable range and using techni-rho and techni-pion masses consistent with the $W+jj$ excess, we find the $\ell^+\ell^-+\met$ rate indeed spans the same region as a $m_H \sim 120-140\,\gev$ SM Higgs. We find a $W+jj$ rate consistent with the CDF excess and good overlap between the technicolor and SM Higgs $\ell^+\ell^-+\met$ rates requires a slightly lighter set of technicolor masses than considered in Ref.~\cite{Eichten:2011sh} and a large $B(\pi_T \rightarrow \tau\nu)$. For the lighter technicolor point, the Higgs-fake signal also has similarly shaped kinematic distributions to a SM Higgs, especially after tight selection and topological cuts are applied. With looser cuts, the differences between a fake and genuine Higgs would be more evident, though perhaps difficult to dig out of larger backgrounds. It would also be interesting to study how Higgs fakes such as technicolor appear when processed through multivariate techniques that are tuned and trained on SM Higgs-like events.

More generally, we introduce a new type of Higgs impostor. This impostor is specific to the $\ell^+\ell^-+\met$ search mode, and is novel in that (at least) one of the leptons comes from a $\tau$ decay rather than from a $W$. The existence of channel-specific Higgs fakes emphasizes that, to verify an excess is indeed the SM Higgs boson, multiple discovery modes must be checked. For example, for the technicolor Higgs fake discussed here, there should be no signal in either the $ZZ(4\ell)$ or $\gamma\gamma$ channels.

\section*{Acknowledgments}

We thank Bogdan Dobrescu, Estia Eichten, Paddy Fox, Roni Harnik, Ken Lane and Graham Kribs for useful conversations. AM is supported by Fermilab operated by Fermi Research Alliance, 
LLC under contract number DE-AC02-07CH11359 with the 
US Department of Energy. AM acknowledges KITP for support while this work was in progress. This research was supported in part by the National Science Foundation under Grant No. NSF PHY05-51164.

\end{document}